\documentclass{mn2e}

\usepackage[dvips]{graphicx}

\renewcommand{\vec}[1]{ {\bmath #1} }

\begin{document}

\title[Chemical feedback in SPH]{Feedback and metal enrichment in 
  cosmological smoothed particle hydrodynamics simulations - I. A model for chemical enrichment}
\author[Scannapieco et al.]{C. Scannapieco, $^{1,2}$\thanks{E-mail: cecilia@iafe.uba.ar (CS);
patricia@iafe.uba.ar (PBT); swhite@mpa-garching.mpg.de (SDMW); volker@mpa-garching.mpg.de (VS)}
P.B. Tissera,$^{1,2}$\footnotemark[1] S.D.M. White,$^{3}$\footnotemark[1] and V. Springel$^{3}$\footnotemark[1] \\
$^1$ Instituto de Astronom\'{\i}a y F\'{\i}sica del Espacio, Casilla de Correos 67,
Suc. 28, 1428, Buenos Aires, Argentina\\
$^2$  Consejo Nacional de Investigaciones Cient\'{\i}ficas
y T\'ecnicas, CONICET, Argentina\\ 
$^3$ Max-Planck Institute for Astrophysics, Karl-Schwarzchild Str. 1, D85748, Garching, Germany}

\date{\today}

\pagerange{\pageref{firstpage}--\pageref{lastpage}}

\maketitle

\label{firstpage}

\begin{abstract}
We discuss a model for treating chemical enrichment by Type II  and Type Ia
supernova (SNII and SNIa) explosions in simulations of cosmological structure formation. Our
model includes metal-dependent radiative cooling and star formation in
dense collapsed gas clumps. 
Metals are returned into the diffuse interstellar
medium by star particles using a local smoothed particle hydrodynamics (SPH) smoothing
kernel. A variety of chemical abundance patterns in enriched gas
arise in our treatment owing to the different yields and lifetimes of
SNII and SNIa progenitor stars. In the case of SNII chemical production, 
we adopt metal-dependent yields.
Because of the sensitive dependence of
cooling rates on metallicity, enrichment of galactic haloes with
metals can in principle significantly alter subsequent gas infall and
the build up of the stellar components. Indeed, in simulations of
isolated galaxies we find that a consistent treatment of
metal-dependent cooling produces  $25$ per cent more stars outside
the central region than simulations  with a primordial
cooling function. In the highly-enriched central regions, the
evolution of baryons is however not affected by metal cooling, because
here the gas is always dense enough to cool.  A similar situation is found in 
cosmological simulations because we include no strong feedback processes
which could spread metals over large distances and mix them into unenriched 
diffuse gas.
We demonstrate
this explicitly with test simulations which adopt suprasolar cooling functions
leading to  large changes  both in the stellar mass and in the metal distributions.
We also find that the impact of metallicity on the star formation histories
of galaxies may depend on their particular evolutionary history. 
Our results hence emphasise the importance of
feedback processes for interpreting the cosmic  metal enrichment.
\end{abstract}


\section{INTRODUCTION}

Over the last decades, our knowledge of the chemical properties of
the Universe and, in particular, of galaxies, has improved
dramatically.  Observations of the Local Universe (e.g. 
Garnett \& Shields 1987; Skillman, Kennicutt \& Hodge 1989; Brodie \& Huchra 1991;
Zaritsky, Kennicutt \& Huchra 1994; Mushotzky et al. 1996; Ettori
et al. 2002; Tremonti et al. 2004; Lamareille et al. 2004) 
as well as at
intermediate and high redshifts (e.g.  Prochaska \& Wolfe 2002; Adelberger et
al. 2003; Kobulnicky et al. 2003;
Lilly, Carollo \& Stockton 2003; Shapley et al. 2004) have resulted in a quite
detailed picture of the chemical history of the stellar populations
and of the interstellar and intergalactic media. It is an important
challenge for the theory of galaxy formation to explain these
observational results in the context of the current cosmological
paradigm. Numerical simulations are an important tool to make specific
predictions for galaxy formation theories and to confront them with
observations, because they can accurately track the 
growth of structure in the dark matter and baryonic components.  It is
hence natural to ask what such simulations predict for the chemical
properties of the universe.

Currently, hydrodynamic cosmological simulations including radiative
cooling and star formation, at best, coarsely reproduce the
observed properties of galaxies. They fail when scrutinised in
detail. Among the most prominent problems are  ``catastrophic
angular momentum loss'' (e.g. Navarro \& Benz 1991; Navarro \& White 1994)
and the difficulty in finding a consistent modelling of supernova (SN) feedback (e.g.
Navarro \& White 1993; Metzler \& Evrard 1994; Yepes et al. 1997; Marri \& White
2003; Springel \& Hernquist 2003, hereafter SH03). The first issue mainly affects
galaxy morphology, size and kinematics, while the latter primarily
influences the efficiency of star formation. Both are the subject of intense
research. Reproducing the chemical properties of observed galaxies in
detail may be viewed as an intertwined third problem.

Regarding the angular momentum problem, recent simulations have shown
some promising advances. Dom\'{\i}nguez-Tenreiro, Tissera \& S\'aiz
(1998) found that the formation of compact stellar bulges can
stabilize the discs and thereby prevent the angular momentum loss
during violent minor merger events, such that disc systems that
resemble observed galaxies are obtained.  Abadi et al. (2003) have
also demonstrated the importance of a dense, slowly rotating
spheroidal component, and pointed out its relevance for comparing
simulated and observed galaxies consistently.  More recently,
Robertson et al. (2004) were able to produce large, stable disc
systems in their cosmological simulations which are comparable in size
to spiral galaxies. In their approach, they introduced a subresolution
model for star formation and feedback which pressurizes the
star-forming gas and stabilizes discs against fragmentation. We note
that all these studies agree on the need for a self-consistent
treatment of SN feedback as a crucial mechanism to regulate star
formation and to reproduce discs similar to observational
counterparts.

SN explosions are thought to play a fundamental role in
the evolution of galaxies since they are considered the most efficient
and ubiquitous mechanism for the ejection of metals and energy into
the interstellar medium (ISM).  Both chemical and energy
feedback can affect the condensation of gas  and consequently the
evolution of galactic systems. On one hand, the presence of metals in
the ISM affects gas cooling times because the radiative cooling rate
depends sensitively on metallicity (Sutherland \& Dopita 1993,
hereafter SD93).  On the other hand, the release of energy is crucial
for regulating star formation through the heating and disruption of
cold gas clouds, and for producing outflows which can transport
enriched material into the intergalactic medium (e.g. Lehnert \& Heckman 1996; Dahlem, Weaver
\& Heckman 1998; Frye,  Broadhurst
\& Ben\'{\i}tez 2002; Rupke, Veilleux \& Sanders 2002;  Martin 2004).  As
suggested by observations and theory, the largest outflows 
(e.g. Larson 1974; White \& Rees 1978; Dekel \& Silk 1986;
White \& Frenk 1991) 
should be able to develop in small systems because of their
shallower potential wells.  Because in hierarchical galaxy formation
scenarios large systems are formed by the aggregation of smaller systems,
energy feedback from SNe is then expected to have important
effects for nearly all systems in the different stages of galaxy formation.

Modelling of chemical feedback  has been addressed already in numerous
studies, mostly with the aim to reproduce the chemical properties of
certain types of galaxies, or of clusters of galaxies (e.g. Larson 1976; 
Tinsley \& Larson 1979; 
Burkert \& Hensler 1988; White \& Frenk 1991; Burkert, Truran \& Hensler 1992; Ferrini et al. 1992;
Theis, Burkert \& Hensler 1992;  Steinmetz \& M\"uller 1994; Kauffmann 1996; Chiappini, Matteucci \& Gratton
1997;  Kauffmann \& Charlot 1998; Boisser \& Prantzos 2000;
Valdarnini 2003).
As first showed by Mosconi et
al. (2001), in order to properly model the
process of metal enrichment in galaxy formation it is necessary to
consider the full cosmological growth of structure in which mergers
and interactions have important effects on the star formation process
and the dynamical evolution of galaxies (see also
Kawata \& Gibson 2003; Tornatore et al. 2004; Okamoto et al. 2005).

Perhaps the most fundamental motivation for the relevance of metal
enrichment for galaxy formation is based on the fact that the cooling
rate of baryons depends on metallicity (SD93).  Radiative cooling in
turn allows the condensation of gas into dense and cold clouds, which
form the reservoir of material available for the formation of
stars. As chemical enrichment is a result of star formation, a
chemical feedback cycle emerges, in which metals can in principle
significantly accelerate the transformation of baryons into
stars. In hierarchical scenarios 
these processes depend significantly on the chemical prescription
adopted (Kaellander \& Hultman 1998; Kay et al. 2000).
These affect galaxy properties such as the luminosity function (White
\& Frenk 1991).
For these reasons, a detailed analysis of the effects of
metal-dependent cooling is clearly important for galaxy
formation studies.
 Substantial numerical challenges in simulations of galaxy formation
come from 
 the large dynamic range required,
 and from 
the multiphase character of the ISM. Gas in very
different physical states  coexists in the ISM, but the standard formulation of 
smoothed particle hydrodynamics (SPH, e.g. Gingold \& Monaghan 1977; Lucy 1977) is not well
suited to deal with this situation. A number of attempts have been
made to solve this problem, ranging from an ad-hoc decoupling of
phases (e.g. Pearce et al. 1999, 2001; Marri \& White 2003) to simple
analytic sub-resolution models (SH03) that work
with an effective equation of state.  

Our new model for SN feedback in SPH cosmological simulations 
has been divided into two stages:
chemical and energy feedback.  In this work we
address the description of the chemical feedback, which
is based on the previous model of Mosconi et al. (2001).  In a
forthcoming paper, we will present the second part of this work: the
implementation of energy feedback in the framework of a multiphase
scheme for the ISM, which extends the approach of Marri \& White (2003). 
While a treatment of chemical enrichment without
energy feedback is an incomplete picture, we think the complexity of the
problem merits a step-wise discussion of our model such that the
primary physical effects can be better understood.

This paper is organized as follows. In Section~\ref{code}, we
summarize the numerical implementation of the chemical model, and in
Section~\ref{isolated-spheres} we analyse the dependence of the
results on numerical resolution and input parameters, using
simulations of isolated galaxies. We then study in
Section~\ref{cosmological} the properties of galaxies formed in
cosmological simulations and examine how they depend on the chemical
model. Finally, in Section~\ref{conclusions} we give our conclusions.

\section{Numerical Implementation}
\label{code}

Our new model for the production and ejection of chemical elements by
SNe  is based in part on the
approach in Mosconi et al.~(2001).  We have implemented our scheme in
the TreePM/SPH code {\small GADGET-2}, an improved version of the
public code {\small GADGET} (Springel, Yoshida \& White 2001) which
manifestly conserves energy and entropy where appropriate (Springel \&
Hernquist 2002).  Note that we do not use the
multiphase treatment and the feedback model developed by SH03 for this code, but we do include
their treatment of ultraviolet (UV) background.  Also, we use a
different parameterisation for star formation, as described in detail
below.

Chemical elements are synthesized in stellar interiors and ejected into
the ISM  by SN explosions.
In order to model the production and distribution of metals in the
context of the simulations, there are three ingredients we have to
consider: the SN rate (i.e. number of exploding stars
per time unit), the  chemical yields (i.e. chemical material
ejected in explosions) and the  typical lifetimes of stars,
which determine the characteristic time of the metal release.

In our model, we include a separate treatment of Type II and Type Ia SNe (SNII
and SNIa). These two types of SNe originate from different
stellar populations, and have different rates, yields and typical
time-scales. We use different yields for SNII and SNIa which, in the
case of SNII, are metal-dependent.
Hence, we will treat SNII and SNIa separately, owing to the need
to model their different characteristics.   For example, SNII produce
most of the chemical elements, except for iron, which is mainly
produced by SNIa.  Another difference between these two types of SN is
that SNII are the endpoint of the evolution of massive stars with
short lifetimes, in contrast to SNIa, which result from the
evolution of binary systems with lifetimes of $\sim$ 1 Gyr.

We assume that initially gas particles have primordial abundances:
  $X_{\rm H}=0.76$ and $Y_{\rm He}=0.24$, and we consider the
  enrichment by the following elements, assuming production 
 only by SNII and SNIa: 
  H, $^4$He, $^{12}$C, $^{16}$O,
  $^{24}$Mg, $^{28}$Si, $^{56}$Fe, $^{14}$N, $^{20}$Ne, $^{32}$S,
  $^{40}$Ca and $^{62}$Zn.  Note that  we do not consider
 nucleosynthesis by intermediate-mass stars. Consequently, we will
 concentrate our analysis on elements such as O or Fe where restricting to
 SN production may be a good approximation.
  In the rest of this Section, we describe
  the most important aspects of the chemical model, namely the star
  formation and cooling prescriptions, and our scheme for the
  production and ejection of metals.

\subsection{Star formation}\label{starformation}

We assume that gas particles are eligible for star formation if they
are denser than a critical value ($\rho > \rho_* = 7.0 \times 10^{-26}
{\rm g \ cm^{-3}}$, where $\rho$ denotes gas density)
and lie in a
convergent flow (${\rm div}\, \vec{v} < 0$)\footnote{See Okamoto et al. (2005) for a different approach.}.
For these particles, we assume a star
formation rate (SFR) per unit volume equal to
\begin{equation}
\dot\rho_\star = c\,\frac{\rho}{\tau_{\rm dyn}},
\end{equation}
where $c$ is a star formation efficiency (we adopt $c=0.1$) and $\tau_{\rm dyn} = 1 / \sqrt{4\pi G\rho}$ 
is the dynamical time of the particle.  We base
the creation of new stellar particles on the stochastic approach of
SH03 (see also Lia, Portinari \& Carraro 2002).  To this end, each gas particle
eligible for star formation in a given time-step is assigned a
probability $p_*$ of forming a star particle given by
\begin{equation}
p_*={m \over{m_*}}\Big[ 1-{\rm exp}\big( -{c\  \Delta t \over {\tau_{\rm dyn}}} \big) \Big],
\end{equation}
where $\Delta t$ is the integration time-step of the code, $m$ is the
 mass of the gas particle, and $m_*$ is defined as $m_*=m_0/N_g$ with
 $m_0$ being the original mass of gas particles at the beginning of
 the simulation and $N_g$ a parameter which determines the number of
 `generations' of stars formed from a given gas particle (we assume
 $N_g=2$).  If a random number drawn from a uniform distribution in
 the unit interval is smaller than $p_*$, we form a new stellar
 particle of mass $m_*$ and reduce the mass of the gas particle
 accordingly. Once the gas particle mass has become smaller than
 $m_0/N_g$, we instead turn it into a star particle.
Newly born stars are assumed to have the same element abundances as
the gas mass from which they form.

The main advantage of this stochastic approach is that all particles
are either purely stellar or purely gas, which avoids the use of
`hybrid' particles with both gaseous and stellar components. The
latter would artificially force the stellar component to evolve dissipately like the gas.  
Note that, as a consequence of
the star formation scheme, the number of baryonic particles in the
simulation does not remain constant. Unlike SH03, we also exchange the
masses of heavy elements explicitly between gas and star particles. As
a result, the mass spectrum of stellar and gas particles is slightly
washed out around otherwise sharp discrete values.

\subsection{Chemical production and distribution}\label{chemical_model}

Our numerical implementation of the enrichment model has three main
components: SNII element production, SNIa element production and metal
distribution.

\vspace{0.3cm}
\noindent {\it SNII element production}  
\vspace{0.3cm}

\noindent In order to estimate the SNII rate
we use a Salpeter Initial Mass Function (IMF) with lower and upper
mass cut-offs of 0.1 and 40 M$_{\sun}$, respectively, and assume that
stars more massive than $8${\rm M}$_{\sun}$ end their lives as
SNII. For the chemical production, we adopt the yields of Woosley \&
Weaver (1995).  These are metal-dependent yields and vary differently
for different elements. We use half of the iron yield of WW95,
 as it is often adopted (e.g. Timmes, Woosley \& Weaver  1995). 
For the sake of simplicity, in most of our
simulations we have assumed that SNII explode (i.e. produce and
eject the chemical elements) within an integration time-step of the
code (which is typically $\leq 10^6$ yr), because SNII originate in
massive stars which have very short characteristic lifetimes of the
order of $10^7$ yr.  However, we have also analysed the effect of
relaxing this instantaneous recycling approximation (IRA) for SNII (see
Section~\ref{SNparameters}).

\vspace{0.3cm}
\noindent {\it SNIa element production } 
\vspace{0.3cm}

\noindent As  a  progenitor model, we
adopt the W7 model of Thielemann, Nomoto \& Hashimoto (1993), which assumes that
SNIa explosions originate from CO white dwarf systems in which mass is
transferred from the secondary to the primary star until the
Chandrasekhar mass is exceeded and an explosion is triggered. It is
generally assumed that the lifetime of such a binary system is in the
range $\tau_{\rm SNIa}= $[$0.1,1\,$] Gyr, depending on the age of the
secondary star (Greggio 1996).  In our model, the material produced
by a SNIa event is ejected when a time $\tau_{\rm SNIa}$ after the
formation of the exploding star has elapsed. We choose this time
randomly within a given range (in most of our experiments we assume
$\tau_{\rm SNIa}=[0.1,1]$ Gyr).  In order to estimate the number of
SNIa, we adopt an observationally motivated range for the relative
ratio of SNII and SNIa rates (e.g. van den Bergh 1991).  We use the
chemical yields of Thieleman et al.~(1993) for the metal production
itself.

\vspace{0.3cm}
\noindent{\it  Metal distribution} 
\vspace{0.3cm}

 We distribute chemical elements ejected
in SN explosions within the gaseous neighbours of exploding star
particles, using the usual SPH kernel interpolation technique
(Mosconi et al. 2001).  Each neighbouring gas particle receives a fraction of the
ejected metals according to its kernel weight.  For this purpose, gas
particle neighbours of stellar particles are identified whenever metal
distribution has to take place, and smoothing lengths for these
stars are calculated by setting a desired number of gas neighbours
that should be enclosed in the smoothing length (we have assumed
the same criterium used for  the smoothing lengths of gas particles).
This is particularly important for
the enrichment by SNIa because of the time delay between the formation
of the stars and the ejection of metals. Note that after a star
particle is created, it is only subject to gravitational forces,
unlike the gas particles, which continue to be subject to
hydrodynamical forces as well.  Consequently, as the system evolves,
star particles can release metals associated with SNIa explosions in a
different environment from where they were born. This, together with
the different lifetimes and yields of SNII explosions, can then
produce non-trivial variations in abundance ratios.

\subsection{Gas cooling}\label{cooling}

\begin{figure}
\includegraphics[width=84mm]{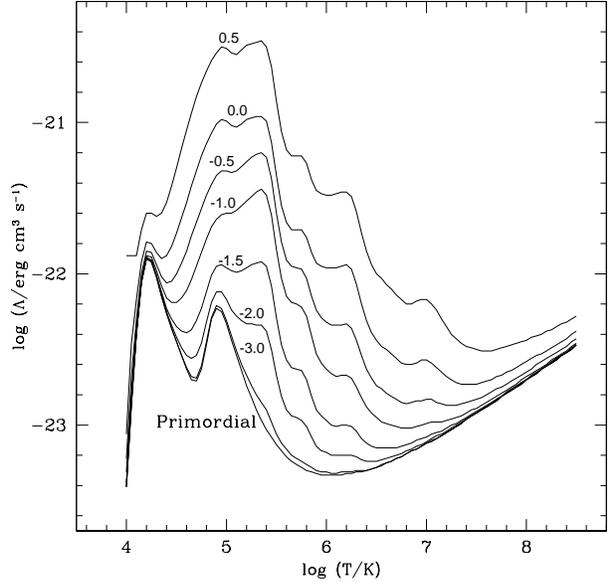}
\caption{Cooling rates for gas of different metallicities, as given by
the [Fe/H] abundance, from suprasolar ([Fe/H]=0.5) to primordial. The
plot is based on data adapted from SD93.}
\label{figcooling}
\end{figure}

It is well known that the cooling rate of thin astrophysical plasmas
depends sensitively on metallicity, in such a way that at a given
density, the cooling rate is higher for a higher metallicity.  In
Fig.~\ref{figcooling} we show the cooling function computed by
SD93 from primordial (i.e. no
heavy elements) to suprasolar abundance ([Fe/H]=$+0.5$).  As one can
see from this figure, the cooling function can show very large
variations with the metal abundance of the gas, depending on the
temperature range considered. For example, the usual estimates of the
cooling time ($\tau_{\rm cool}$) for a gas particle at the critical
density ($\rho_{*}$) and at a temperature of $ T_{*}= 4\times 10^4$ K
yield that $\tau_{\rm cool}$ is 50 times larger for primordial gas
than for a mildly suprasolar medium ([Fe/H]$=0.5$), 18 times larger
than that of a solar abundance one, and still twice as large as the
corresponding time for gas with [Fe/H]$=-1.5$. The temperature range
of [$10^5-10^6$] K shows the largest differences, right in the
temperature range relevant to the abundant population of dwarf
galaxies.

Including effects owing to metal line cooling is therefore crucial for
realistic modelling of galaxy formation because it affects directly
the relation between cooling time and dynamical time of forming
galaxies.  We therefore include a treatment of radiative cooling
consistent with the metal content of the gas component produced by the
enrichment model. We use the tables computed by SD93 for the
metal-dependent cooling functions, and interpolate from them for other
metallicities as needed. For particles with iron abundance larger than
$0.5$, we adopt the suprasolar cooling rate (i.e. that for
[Fe/H]$>+0.5$). In cosmological simulations, we also include a
redshift-dependent photo-heating UV background, with an intensity
evolution that follows the model of Haardt \& Madau (1996).

\section{Results for isolated  galaxy models}
\label{isolated-spheres}

As first tests of our model, we consider simulations of isolated haloes
set up to form disc galaxies at their centre. This allows us to
examine the dependence of the results on resolution and on different
parameters of the chemical enrichment model. The initial conditions
consist of a static dark matter potential corresponding to a Navarro,
Frenk \& White (1996, 1997)  profile of concentration $c = 20$, and a baryonic gas phase
initially in hydrostatic equilibrium in this potential.  Our typical
system has virial mass $M_{\rm 200}=10^{12}\, h^{-1} M_{\odot}$
($h=0.7$), $10$ per cent  of which is in the form of baryons ($M_{\rm
bar}=10^{11} h^{-1} \ M_{\odot}$).  The initial radius of the system
is $r_{200}= 160 \, h^{-1}{\rm kpc}$ and the gas has an initial angular
momentum characterized by a spin parameter $\lambda=0.1$.
We have followed the evolution of this system for $1.5\, \tau_{\rm dyn}$, where
$\tau_{\rm dyn}= 7$ Gyr is the dynamical time at $r_{200}$.

These idealized initial conditions yield a  simple model for disc
formation, which is an ideal test bench for the performance and
validity of the code, and the dependence of the results on the free
parameters. We stress however that these models are not meant to
provide a realistic scenario for the whole galaxy formation process.
Results for full cosmological simulations will be discussed in Section
4, where we analyse the properties of galactic objects formed in
$\Lambda$ cold dark matter (CDM) models with virial masses of $M_{\rm vir} \approx
10^{12}\, h^{-1} M_\odot$, comparing them also to the results of the
idealized tests discussed here.

\subsection{Numerical resolution}

Both the star formation process and the distribution of metals can be
affected by numerical resolution. Particularly important for our model
is the fact that the smoothing length of the particles (i.e. the
radius which encloses $\approx 32$ gaseous neighbours) determines the
region where the newly released chemical elements are injected.
Mosconi et al.~(2001) found that chemical abundances could be
oversmoothed in low resolution simulations because  metals are
distributed over a comparatively large volume and, consequently, are
more efficiently mixed with surrounding gas.  On the other hand,
high numerical resolution could also produce spurious results if the
metal mixing mechanism becomes too inefficient. Provided
other physical mixing mechanisms are not operating, the simulation
results might show an artificially large dispersion in  gas
metallicities.

In order to test the effects of numerical resolution, we carried out
simulations of the isolated disc system described above using three
different initial numbers of gas particles: 2500 (R1), 10000 (R2) and
40000 (R3), with mass resolutions of $4\times 10^7$,
$10^7$ and $0.25\times 10^7\, h^{-1} M_{\odot}$ for
R1, R2 and R3, respectively. We have set the gravitational softening 
to $0.4\ h^{-1}$ kpc. In these three runs we use
metal-dependent cooling functions for the gas particles according to
the tables by SD93.  In Table~\ref{simulations} we summarize the main
characteristics of these simulations.

\begin{figure}
\includegraphics[height=45mm]{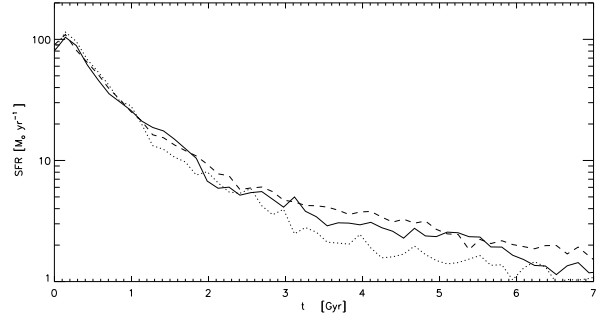}
\caption{Evolution of the SFR for our test simulations
of isolated disc galaxies, as a function of different numerical
resolution. R1 (dotted line) has 2500 particles, R2 (solid line) 10000
particles, and R3 (dashed line) 40000 particles.}
\label{sfr-resolution}
\end{figure}

In Fig.~\ref{sfr-resolution}, we show the evolution of the SFRs
 obtained for R1 (dotted line), R2 (solid line) and R3
(dashed line).  Most of the stellar mass ($92$, $88$ and $86$ per cent in
R1, R2 and R3, respectively) is formed at times $t<3$ Gyr ($43$ per cent of
the dynamical time) in the three runs. Only at later times, we find
that the lowest resolution test simulation (R1) yields a significantly
lower result for the SFR, indicating that the number of gas particles
has dropped so much by the conversion into stars that discrete effects
become appreciable.  However, the final stellar mass fractions (i.e.
stellar mass formed divided by the total baryonic mass) are very
similar in R1, R2 and R3.

\begin{figure*}
\includegraphics[height=120mm]{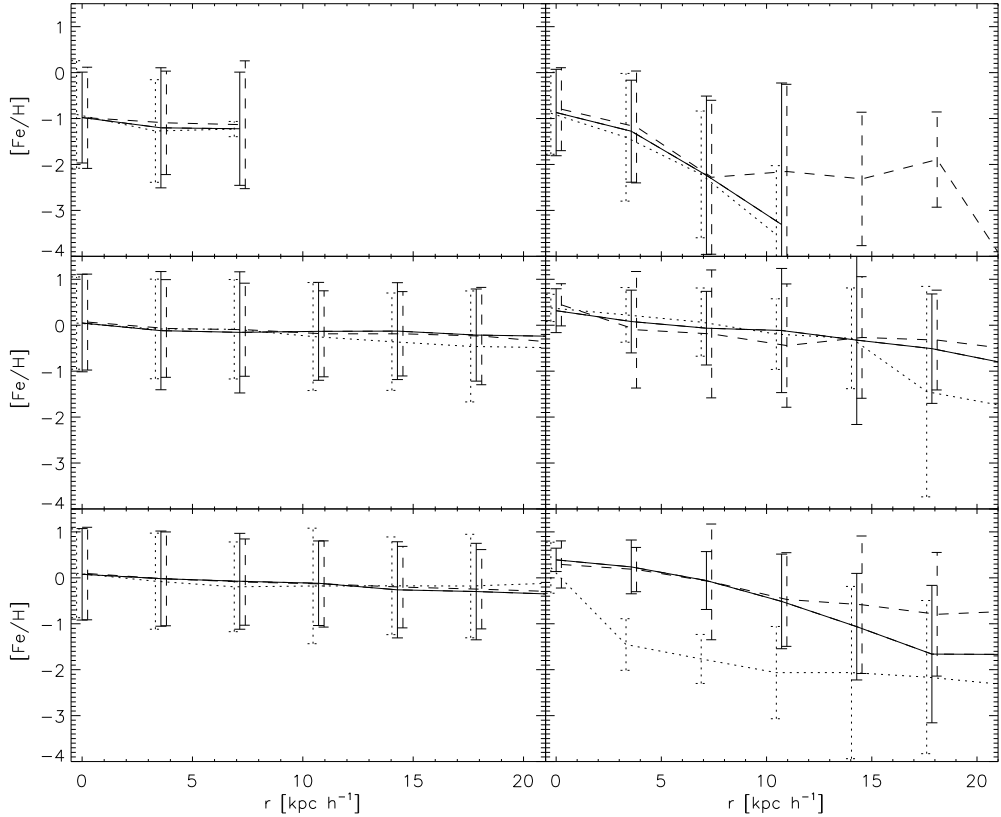}
\caption{Metallicity profiles for the stellar (left panels) and
gaseous (right panels) components of the idealized disc tests
performed with different resolution: R1 (dotted line, 2500 particles),
R2 (solid line, 10000 particles) and R3 (dashed line, 40000
particles). The different rows show different times of the evolution:
$t=0.13$ Gyr (upper panels), $t=2.5$ Gyr (middle panels) and $t=7$ Gyr
(lower panels). The error bars correspond to the rms scatter 
 of [Fe/H] around the mean in each radial bin.}
\label{perfiles-resolution}
\end{figure*}

This can be also appreciated from Fig.~\ref{perfiles-resolution},
where we show the mass-weighted metallicity profiles for R1 (dotted
line), R2 (solid line) and R3 (dashed line), in the inner region and
for the stellar (left column) and gaseous (right column) components
and for three different times, $t = 0.13$ Gyr (upper panels), $t =
2.5$ Gyr (middle panels) and $t = 7$ Gyr (lower panels).  We can see
that the stellar abundance profiles are insensitive to resolution. The
mean iron abundance differs by less than $0.03$ dex between the
different simulations.  Recall that the stellar metallicity of newly
forming stars is given by the local gas metallicity. This is directly
visible at the beginning of the simulation where the gaseous and
stellar components show a similar level of enrichment in star-forming
regions (upper panels in Fig.~\ref{perfiles-resolution}), while at
later stages of the evolution (middle and lower panels) this
constraint becomes hidden as a result of the superposition of stars of
different ages.

However, as  can be seen from the right panels of
Fig.~\ref{perfiles-resolution}, the mass-weighted metallicity profiles
for the gas components show
stronger differences with resolution than their stellar counterparts.  
At the beginning of the
evolution, the profiles of the three test simulations are still very
similar, although the enriched gas is more spread in the highest-resolution
test.  However, the gas profiles differ more at later times
when factors such as variations in the consumption time of the gas and
differences in the spatial distribution of stars begin to affect the
radial distribution of metals.  At the end of the simulations (lower
right panel), the mean gaseous iron abundance for R1 is significantly
lower than those for the other two simulations.  This decrease in the
metallicity is produced by the infall of less enriched gas. This gas is
not becoming dense enough to trigger star formation activity at the
level of R2 and R3, and the infall tends to dilute the metal content in the
ISM.  Hence, the fact that the gas component in R1
shows a lower iron abundance is driven by its poorer gas resolution.
Note that as a result of the consumption of most of the gas, the final
number of gas particles in R1 is very low ($\sim 60$ in the inner $30$
kpc) and, as a consequence, the distribution of metals in the gas
component is strongly affected by numerical noise. On the other hand,
the results for the tests with 10000 and 40000 particles 
converge quite well both in their stellar and gaseous chemical
properties. Note that the rms scatter of particles
within each radial bin is similar for the different tests
indicating that, on average, the metal mixing is not significantly affected
by resolution if we consider at least an initial number of particles of $2500$ or more.

\begin{table*}
\center
\caption{Chemical model parameters for the different tests of the
idealized isolated galaxy: initial number of gas particles $N_{\rm
gas}$, typical lifetimes associated with SNII ($\tau_{\rm SNII}$) and SNIa
($\tau_{\rm SNIa}$) in Gyr, rate of SNIa and cooling functions adopted.}
\begin{tabular}{lrcccl }\hline
Test &  $N_{\rm gas}$ & $\tau_{\rm SNII}$& $\tau_{\rm SNIa}$ & Rate$_{\rm SNIa}$ & Cooling Function\\\hline
R1    &  2500  & 0  & $0.1-1$         & 0.0015       &Metal-dependent\\
R2    & 10000  & 0 & $0.1-1$         & 0.0015       &Metal-dependent\\
R3    & 40000  & 0 & $0.1-1$         & 0.0015       &Metal-dependent\\\hline
P1    & 10000  & 0 & $0.1-1$         & 0.0033       &Metal-dependent\\
P2    & 10000  & 0 &   $0$         & 0.0015    &Metal-dependent\\\hline
P3    & 10000  & see Section~\ref{SNparameters} & $0.1-1$         & 0.0015       &Metal-dependent\\\hline
PC    & 10000  & 0 & $0.1-1$         & 0.0015       &Primordial\\
SC    & 10000  & 0 & $0.1-1$         & 0.0015       &Suprasolar\\\hline
\end{tabular}
\label{simulations}
\end{table*}

\subsection{Dependence on assumed supernova characteristics}\label{SNparameters}

In this Section we analyse the dependence of our results on important
input parameters of the chemical model, namely the 
IRA for SNII, and the rate and lifetimes
associated with SNIa.  For this purpose we run three additional test
simulations. In runs P1 and P2, we change the SNIa parameters compared
with the simulation R2, which we take as a fiducial reference model,
while in run P3 we release the IRA condition for SNII.  The main
characteristics of these simulations are listed in
Table~\ref{simulations}.

\begin{figure}
\includegraphics[height=45mm]{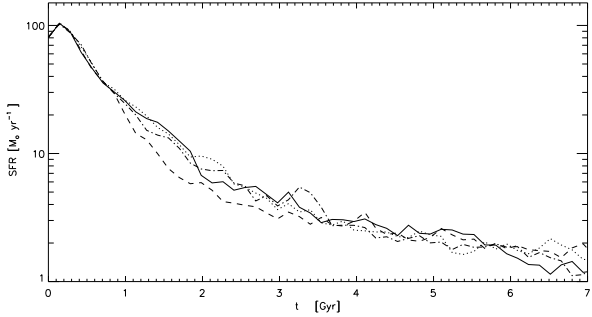}
\caption{SFR for the experiments of the idealized disc
galaxy run with different SN parameters: R2 (solid line, standard
test), P1 (dotted line, higher SNIa rate), P2 (dashed line, IRA
for SNIa) and P3 (dashed-dotted line, relaxing the
IRA for SNII). }
\label{sfr-diffSN}
\end{figure}

\begin{figure}
\includegraphics[height=45mm]{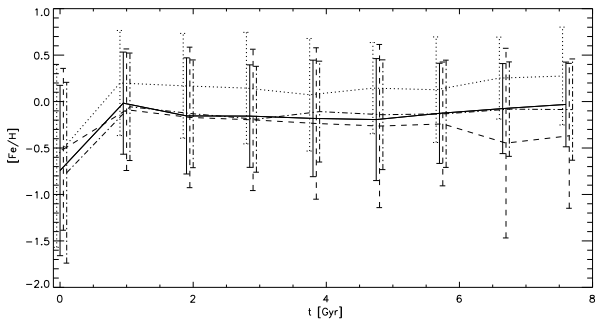}
\caption{AMR for the stellar component in the
tests of the idealized disc galaxy run with different SN parameters: R2
(solid line, standard test), P1 (dotted line, higher SNIa rate), P2
(dashed line, IRA
for SNIa), and P3 (dashed-dotted line,
relaxing the IRA for SNII).  The
error bars correspond to the rms scatter of [Fe/H] around the mean. }
\label{amr-diffSN}
\end{figure}

\begin{figure}
\includegraphics[height=45mm]{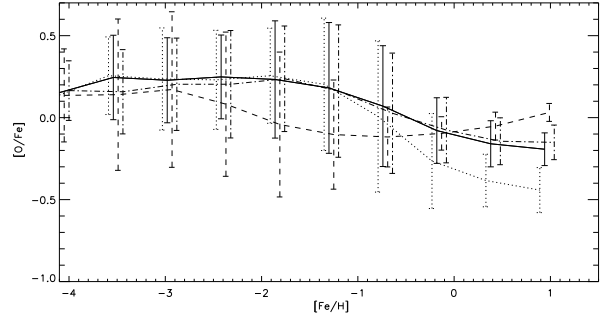}
\caption{[O/Fe] abundance as a function of [Fe/H] for the stellar
component in experiments of the idealized disc galaxy run with
different SN parameters: R2 (solid line, standard test), P1 (dotted
line, higher SNIa rate), P2 (dashed line, IRA
for SNIa), and
P3 (dashed-dotted line, relaxing the IRA for SNII).  The error bars correspond to the rms scatter
of [O/Fe] around the mean. }
\label{ofeh-stars-diffSN}
\end{figure}

In Fig.~\ref{sfr-diffSN}, we compare the evolution of the SFR
 of the standard R2 run (solid line) with P1 (dotted line), in which the
SNIa rate was increased.  We find no significant differences between the SFRs
of these tests.
However, we find a larger difference in their
stellar age-metallicity relations (AMR), as seen in
Fig.~\ref{amr-diffSN}, although the results appear consistent within the
standard deviation based on counting statistics.  In this plot, an increase in
the SNIa rate translates into a shift towards larger values of stellar
metallicities, since the ISM is more strongly enriched by
heavy elements. A change of the SNIa rate also impacts other chemical properties of the
systems.  For example, in Fig.~\ref{ofeh-stars-diffSN} we show the [O/Fe]
abundance as a function of [Fe/H] for stars in R2  and P1 at
$\tau_{\rm dyn}$. The main difference is found for the high-metallicity
material ([Fe/H] $>$ -0.5) which shows in P1 a lower $\alpha$-enhancement compared
to that in R2, as expected.

We have also tested the sensitivity of the results on the lifetime of
binary systems associated with SNIa. The importance of including SNIa
has been previously pointed out by Greggio \& Renzini (1983) and
Mosconi et al. (2001), among others.  Motivated by these previous results, we carried
out a simulation P2 where we assumed an IRA for SNIa explosions
(see Table~\ref{simulations}).  
The main effect of instantaneously releasing  the chemical production 
of SNIa can be appreciated from 
Fig.~\ref{ofeh-stars-diffSN}, which shows the mean $\alpha$-enhancement
of the simulated stellar population. These abundance ratios  exhibit
a behaviour which is in open disagreement with observations (Pagel
1997).
 Note, however, that these observational results correspond to solar
neighbour stars, while the simulated ones include information from  the whole stellar
component.

Finally, in order to analyse the effects of the IRA 
for SNII, we tested a more detailed
calculation of the lifetime of massive stars. To this end, we
convolved the IMF with the lifetime of massive stars at different
intervals of metallicity (Raiteri, Villata \& Navarro 1996).  In this
way, we obtained a range of mean lifetimes varying with stellar
metallicity from $6.3\times 10^6$yr for $Z=5\times 10^{-4} Z_\odot$ to
$1.6\times 10^7$yr for $Z=Z_\odot$, where $Z$ denotes mean stellar
metallicity and $Z_\odot$ is the solar metallicity.  We run a test of
our idealized initial condition with the same chemical parameters used
in R2 but including these metallicity-dependent lifetimes for SNII
(P3, see Table~\ref{simulations}).  We found no significant
differences between the two tests, either in the SFR or in the iron
abundance, as can be seen from Figs.~\ref{sfr-diffSN}
and~\ref{amr-diffSN}.  The similar behaviour of these two runs is also
found in other chemical properties of the systems, such as the
relation shown in Fig.~\ref{ofeh-stars-diffSN}. Recall that a drawback
of a detailed description of chemical enrichment can be its high
computational cost, so an important practical goal is to keep the
model sufficiently simple to allow its use in large-scale cosmological
simulations. While in this context the IRA for SNIa  in general cannot
reproduce certain observed trends, for SNII it appears to be a simple
and sufficiently accurate simplification, at least for these simple
tests that do not include energy feedback.

\subsection{Effects of metal-dependent cooling}
\label{cooling}

\begin{figure}
\includegraphics[height=45mm]{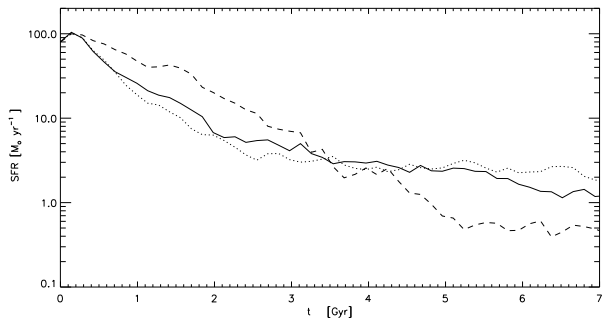}
\caption{SFR for the idealized disc test run with
suprasolar (SC, dashed line), primordial (PC, dotted line) and
metal-dependent (R2, solid line) cooling functions.}
\label{sfr-diffCool}
\end{figure}

Finally, we performed two more runs of our idealized isolated galaxy
in order to highlight effects owing to a metal-dependent cooling
function.  In these fiducial tests, we assumed a constant metallicity,
either primordial (PC) or suprasolar (SC, [Fe/H]$=+0.5$) for the
cooling function. The parameters for star formation and SN rates were
the same as in the standard simulation R2 discussed earlier (see
Table~\ref{simulations}).

\begin{figure*}
\includegraphics[height=200mm]{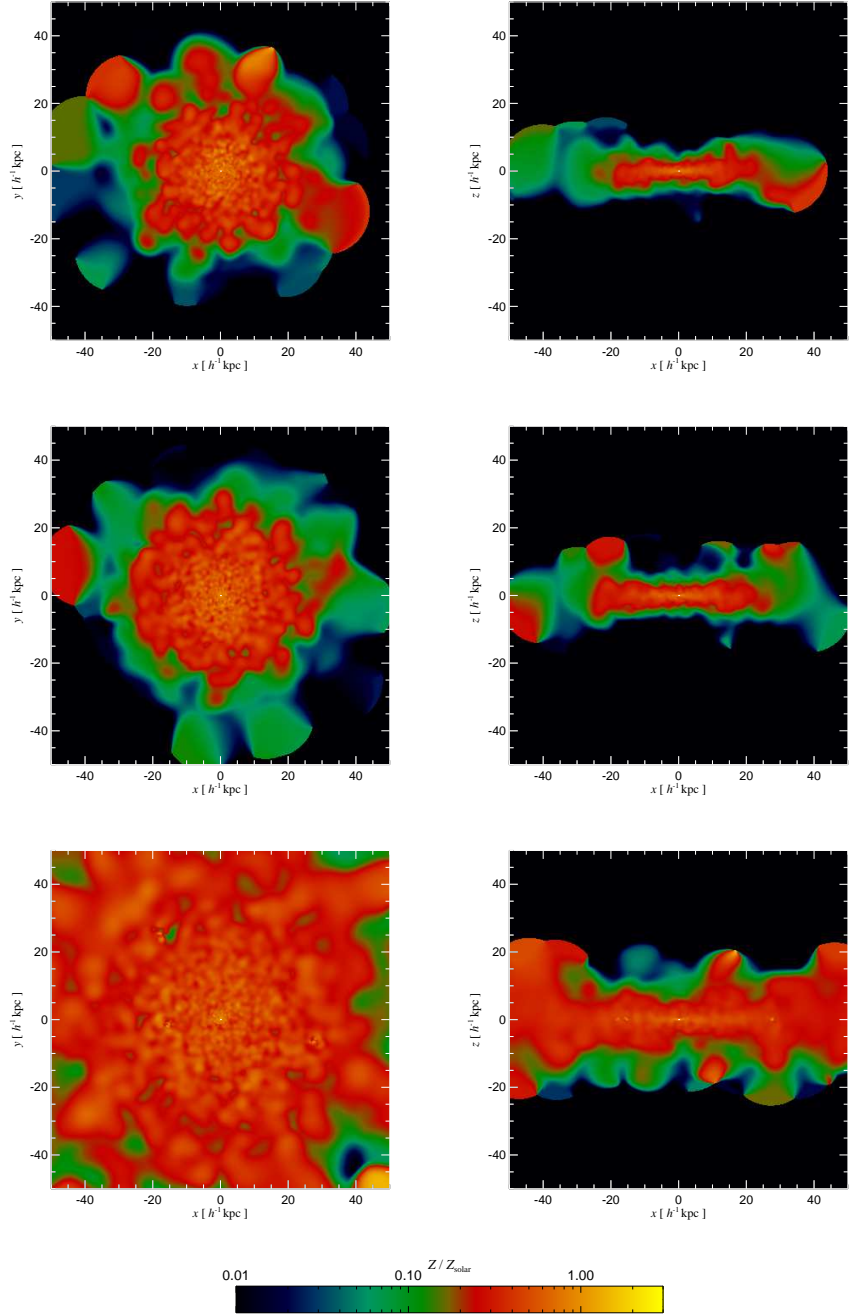}
\caption{Face-on (left-hand panels) and edge-on (right-hand panels)
surface metallicity distribution for the stellar component of the
isolated disc galaxy tests run with primordial (PC, upper row),
metal-dependent (R2, middle row) and suprasolar (SC, lower row)
cooling rate functions. The metallicity scale is also shown.}
\label{mapa}
\end{figure*}

In Fig.~\ref{sfr-diffCool}, we compare the evolution of the SFR
 for R2 (solid line), PC (dotted line) and SC (dashed
line).  In the early stages ($t \leq 4$ Gyr), the test run with
suprasolar abundance cooling function (SC) results in a SFR
 higher by up to a factor of $3$ compared with the
simulation R2 with self-consistent cooling function, while the run PC
with primordial cooling lies lower by up to a factor $1.5$.
After the first $4$ Gyr, the SFR in the SC run decreases
significantly, indicating that most of the gas has already been
consumed and transformed into stars. Likewise, after this period,
simulation PC has the highest residual level of SFR as a result of the
larger amount of gas left over and available for star formation.

\begin{figure}
\includegraphics[height=45mm]{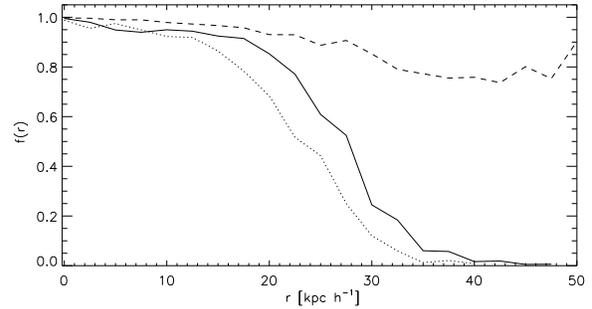}
\caption{  Fraction of baryons in stellar form
 $f(r)$  for the idealized disc tests
 run with suprasolar  (SC, dashed line),
primordial (PC, dotted line) and metal-dependent (R2, solid line) cooling functions.}
\label{perfil-mstar}
\end{figure}

\begin{figure}
\includegraphics[height=82mm]{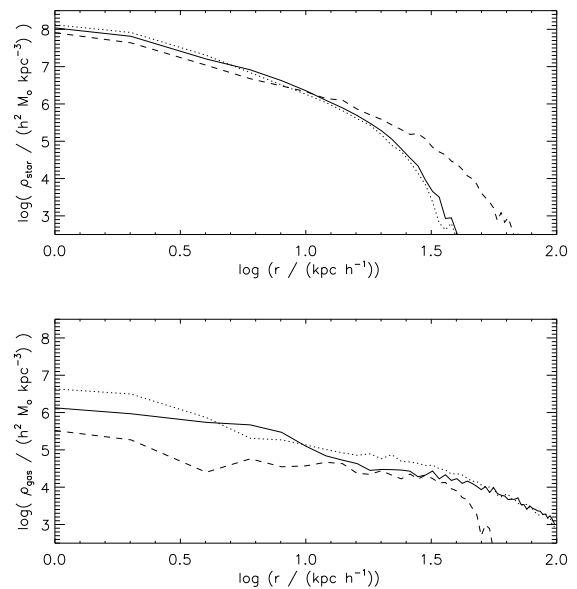}
\caption{ Density profiles for the stellar (upper panel) and gaseous
(lower panel) components for the idealized disc tests
 run with suprasolar  (SC, dashed line),
primordial (PC, dotted line) and metal-dependent (R2, solid line) cooling functions.}
\label{perfil-density-star-gas}
\end{figure}

As a result of the differences in the level of star formation activity
among the runs with different cooling functions, the metallicity
distribution is also affected.  In Fig.~\ref{mapa} we show metallicity
maps for the stellar components corresponding to face-on (left panels)
and edge-on (right panels) projections of the discs, for the PC (upper
panel), R2 (middle panel) and SC (lower panel) simulations.  Comparing
the upper to the lower panels, the distribution of stars is more
extended and the level of enrichment is higher in the SC run.  In
fact, the increase in the cooling efficiency produces an increase in
the star formation activity at larger radius, increasing the amount of
metals and affecting their relative distribution.  This can also be
appreciated from Fig.~\ref{perfil-mstar}, which shows the fraction of
baryons in stellar form $f(r)$ as a function of radius for tests R2
(solid line), PC (dotted line) and SC (dashed line) at $\tau_{\rm
dyn}$. These fractions were calculated by summing up the stellar mass
in radial bins, normalized to the total baryonic mass within the
corresponding bin.  In particular, at $20\ h^{-1}$ kpc the
distribution functions for R2 and SC are $25$ and $35$ per cent higher than
that for the PC run, respectively.  Note that the impact of using
different cooling functions becomes more important as we go to outer
regions where, in the limiting case of SC, the high cooling efficiency
leads to a substantial star formation activity even outside $30\
h^{-1}$ kpc. The behaviour found in this figure is present already in
the early stages of the evolution. The efficient transformation of gas
into stars at larger radius also limits the supply of baryons to inner
regions. Hence, the whole mass distribution is affected by the use of
different cooling functions. This is clear from
Fig.~\ref{perfil-density-star-gas}, where we show the density profiles
of the stellar (upper panel) and gaseous (lower panel) components for
these tests.

\begin{figure*}
\includegraphics[height=120mm]{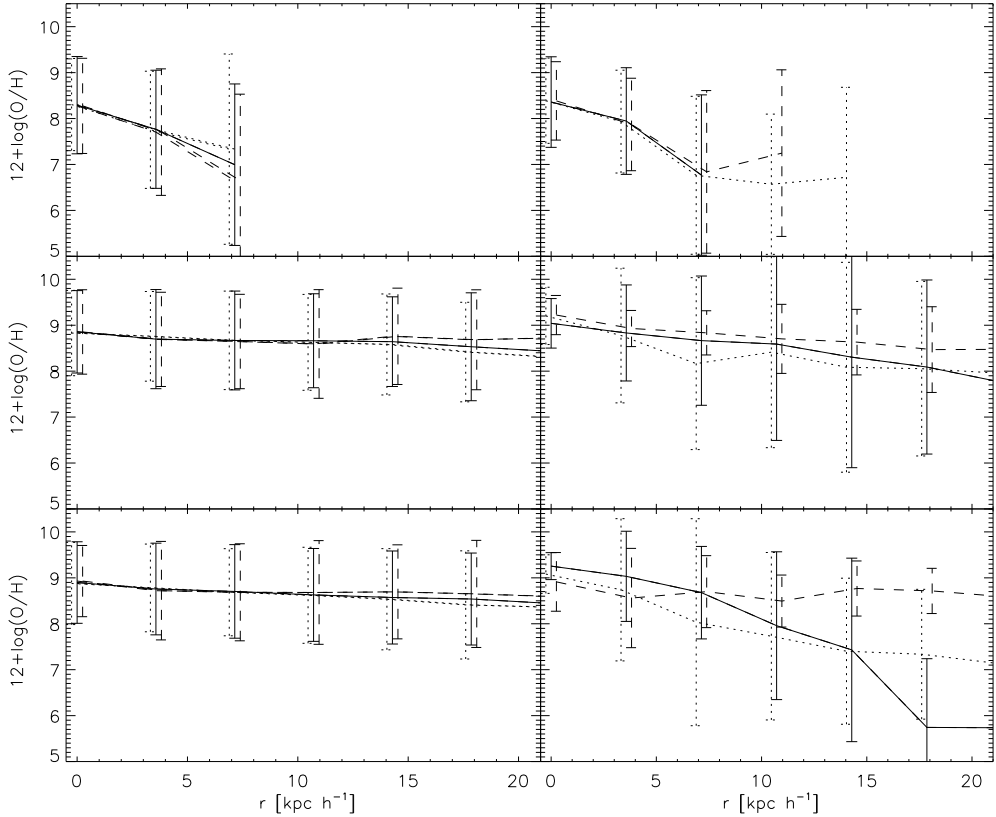}
\caption{Oxygen profiles for the stellar (left panel) and gaseous
(right panel) components in experiments of the idealized disc galaxy
run with different cooling functions: R2 (solid line, metal-dependent
cooling), PC (dotted line, cooling function for primordial abundance)
and SC (dashed line, cooling function for suprasolar abundance).  The
error bars correspond to the rms scatter of 12 +log (O/H) around
the mean.}
\label{perfil-gas-diffCool}
\end{figure*}

Finally, in Fig.~\ref{perfil-gas-diffCool} we show the mass-weighted oxygen profiles
for the stellar (left panels) and gaseous (right panels) components
for the tests PC (dotted line), R2 (solid line) and SC (dashed line),
at three different times: $t = 0.13$ Gyr (upper panels), $t = 2.5$ Gyr
(middle panels), $t = 7$ Gyr (lower panels).  We find no significant
differences in the abundance distributions for the stars in these
tests at any time.  However, the metallicity of the gaseous component
is more sensitive to the adopted cooling functions.  At the beginning
of the simulations, the results for the three runs are similar but
after the initial collapse (middle panel) the level of enrichment
increases from PC to R2 and SC, even though the differences remain
quite small overall.  At $\tau_{\rm dyn}$ (lower panel), both the PC
and R2 simulations show decreasing metallicity profiles, conversely to
the case of SC which shows a flat relation. Here metals are more
uniformly distributed and the ISM is enriched out to larger radii.
Recall that the gas metallicities reflect an instantaneous state of
the chemical properties of the system, while the stellar population
gives an integrated account of the history of the chemical properties
of the galaxy.

In summary, the results of this section show that a detailed study of
the metallicity properties of galaxies requires a self-consistent
treatment of radiative cooling that accounts for metal line
cooling. The resulting effects will, however,  depend strongly on how
 heavy elements are spread into diffuse unenriched gas.

\section{Results for Cosmological Simulations}
\label{cosmological}

We here discuss first results for full cosmological simulations of the
hierarchical growth of galaxies, allowing us to more realistically
assess effects of chemical enrichment on galaxy properties.  To this
end, we have run a cosmological simulation (C2) in a CDM
 universe with cosmological parameters $\Omega_{\rm m}=0.3$,
$\Omega_{\rm \Lambda}=0.7$, $\Omega_{\rm b}=0.04$, $H_{\rm 0}=100
h^{-1} {\rm km \ s}^{-1} {\rm Mpc}^{-1}$, $h=0.7$, and
$\sigma_8=0.9$. We used a periodic simulation box of 10 $h^{-1}$ Mpc
on a side, populated initially with $N=2\times 80^3$ particles,  
which yields an initial mass resolution of $\approx 1.4\times 10^8$ and
$\approx 2.16\times 10^7$ $h^{-1}$ M$_\odot$
for dark matter and gas particles, respectively. We have
adopted a maximum gravitational softening of $5$ $h^{-1}$ kpc for dark
matter, gas and star particles. 
We used a metal-dependent, self-consistent cooling function for the gas
component.  
Table~\ref{simulations2} summarizes the main
characteristics of this run.

In order to test the dependence of our results on mixing processes and
metal-dependent cooling, we have also carried out two additional runs
with the same parameters, but a constant metallicity for the cooling
function. As a limiting case of a very enriched medium, in
simulation C3  we used
suprasolar abundance (see Table ~\ref{simulations2}),
while in simulation C1 we took primordial cooling in order to model a
situation where mixing processes are completely absent. These two
cases should hence bracket reality.

\begin{table*}
\center
\caption{Chemical model parameters for the cosmological tests:
initial number of gas and dark matter particles
$N_{\rm gas}$ and $N_{\rm DM}$, typical lifetimes associated with 
SNII ($\tau_{\rm SNII}$) and SNIa ($\tau_{\rm SNIa}$) in Gyr,
rate of SNIa and cooling functions adopted.}
\begin{tabular}{lrcccl }\hline
Test &  $N_{\rm gas}= N_{\rm DM}$ & $\tau_{\rm SNII}$& $\tau_{\rm SNIa}$  & Rate$_{\rm SNIa}$ & Cooling Function\\\hline
C1    & 512000  & 0 & $0.1-5$         & 0.0015       &Primordial\\
C2    & 512000  & 0 & $0.1-5$         & 0.0015       &Metal-dependent\\
C3    & 512000  & 0 & $0.1-5$         & 0.0015       &suprasolar\\\hline
\end{tabular}
\label{simulations2}
\end{table*}

In our cosmological simulations, we identified galactic objects  at $z=0$ as
virialized structures using a density-contrast criterion with
overdensity $\delta \rho /\rho \approx 178\ \Omega_0^{-0.6}$ (White,
Efstathiou \& Frenk 1993). In our analysis we focus on six haloes with
masses similar to the idealized disc test R2 ($\simeq 10^{12} {\rm
M}_{\odot}$). They are resolved with a similar or higher number of
particles ($N \ge 10000$, see  Table~\ref{N-glos}). Note that as
gas/star particles do not have the same mass, the number of particles in these
components does not, however, trace the mass.
The main properties of these objects are
given in Table~\ref{glos}.

\begin{figure*}
\includegraphics[height=120mm]{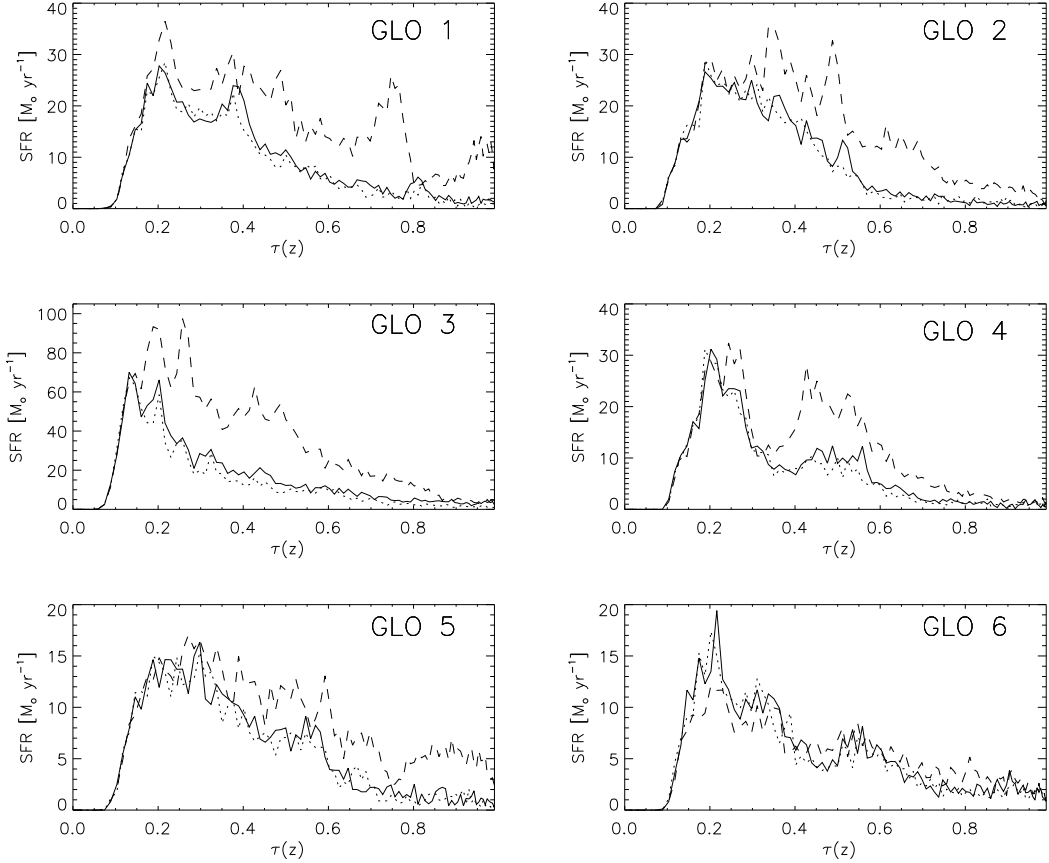}
\caption{SFR for the six selected galactic objects from the
cosmological simulations run with primordial (dotted line),  metal-dependent (solid line) and
suprasolar (dashed lines) cooling functions.}
\label{sfr-glos}
\end{figure*}

\begin{table*}
\center
\caption{Number of dark matter (${N}_{\rm DM}$), gas (${N}_{\rm gas}$), and
stars (${N}_{\rm star}$) within the virial radius   for the
six galactic objects (GLOs) identified in the cosmological tests run with
primordial (C1), metal-dependent (C2) and suprasolar (C3) cooling functions.}

\begin{tabular}{cccccccccc}\hline
    &  & C1 &    & &   C2  & &    &  C3  & \\\hline 
GLO &   $N_{\rm DM}$ & $N_{\rm gas}$ & $N_{\rm star}$ &  $N_{\rm
DM}$ & $N_{\rm gas}$ & $N_{\rm star}$ & $N_{\rm DM}$ & $N_{\rm gas}$ & $N_{\rm star}$\\\hline
1   &  8591 & 2149 & 12838 & 8552 & 1881 & 13382 & 8739 & 1486 & 19368  \\
2   &  5454 & 1337 & 7573  & 5482 & 1270 & 8013  & 5684 & 1431 & 12313  \\
3   &  21816 & 8947 & 24218& 21790 & 6930 & 27105& 22275 & 2442 & 43430  \\
4   &  5738 & 1858 & 7785  & 5656 & 1740 & 8240  & 5840 & 1569 & 11956  \\
5   &  8738 & 2724 & 12385 & 8688 & 2451 & 13362 & 8934 & 1736 & 20465  \\
6   &  4980 & 1506 & 7868  & 4966 & 1476 & 8166  & 5150 & 1512 & 10482  \\\hline
\end{tabular}
\label{N-glos}   
\end{table*}

\begin{table*}
\center
\caption{Virial  mass ($M_{\rm vir}$), stellar mass ($M_{\rm star}$,
both in $10^{10} h^{-1} M_\odot$) and
optical radius ($r_{\rm opt}$ in $h^{-1}$ kpc) for the
six galactic objects  (GLOs) identified in the cosmological tests run with
primordial (C1), metal-dependent (C2) and suprasolar (C3) cooling functions.}

\begin{tabular}{cccccccccc}\hline
    &  & C1 &  &   &   C2  & & &    C3  & \\\hline 
GLO &  $M_{\rm vir}$ &$M_{\rm star}$& $r_{\rm opt}$& $M_{\rm vir}$ &$M_{\rm star}$& $r_{\rm opt}$ &  $M_{\rm vir}$ &$M_{\rm star}$& $r_{\rm opt}$ \\\hline
1   &  139.28 & 13.74 & 15.05 &  138.81 & 14.35 & 14.20 & 146.70 & 20.72 & 15.70  \\
2   &  87.79  & 8.11  & 15.38 &  88.52  & 8.58  & 14.15 &  95.98 & 13.16 & 14.99  \\
3   &  352.44 & 25.91 & 13.31 &  351.07 & 29.06 & 12.51& 365.50 & 46.56 & 12.40  \\
4   &  93.04  & 8.32  & 14.30 &  92.09  & 8.79  & 13.47 &  98.26 & 12.80 & 12.73  \\
5   &  142.10 & 13.26 & 12.06 &  141.79 & 14.30 & 11.78 & 151.09 & 21.89 & 10.90  \\
6   &  81.63  & 8.40  & 12.38 &  81.70  & 8.72  & 12.58 &  86.71 & 11.20 & 14.71  \\\hline
\end{tabular}
\label{glos}   
\end{table*}

In Fig.~\ref{sfr-glos}, we compare the star formation histories for
the selected galactic objects in C1 (dotted line), C2 (solid line) and C3 (dashed line), as a
function of look-back time, i.e. $\tau(z)=1-[1+z]^{-3/2}$.  Note
that the global features of the SFRs are preserved in the three runs,
although the detailed levels of activity are different.  In general,
the SFRs of the galactic objects formed in C3 are systematically higher than
the corresponding ones in C2, as expected.  Note that because galactic objects
formed in C3 have higher SFRs than those in C2, their final stellar masses
are much larger than the corresponding ones in C2
(see Table~\ref{glos}). If we compare the SFRs obtained for C1 and C2, 
although the differences are small, the SFRs in C1 are systematically
lower than those in C2.
It is noteworthy that the impact of metallicity on
the star formation history seems to depend on the particular evolutionary
histories of the systems, as can be appreciated from Fig.~\ref{sfr-glos}.

\begin{figure}
\includegraphics[height=45mm]{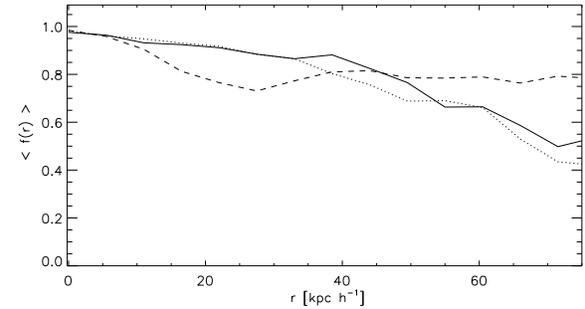}
\caption{Mean  stellar mass fraction as a function of radius for the six
selected objects in the cosmological simulations run with primordial (C1, dotted line), 
metal-dependent (C2, solid line) and suprasolar (C3, dashed line) cooling
functions as a function of radius.}
\label{mean-perfiles-mstar-obj}
\end{figure}

In Fig.~\ref{mean-perfiles-mstar-obj}, we show the mean stellar mass fraction 
 $\left< f(r) \right>$ as a function of radius,
estimated from the six selected galactic objects in C1 (dotted line), C2 (solid line) and C3
(dashed line).  
As in Section~\ref{cooling}, the
distribution function for each galactic object was calculated by summing up the stellar mass in
bins of radius normalized to the total baryonic mass within the corresponding bin.
As we found for the tests of the isolated galaxies,
there are clear differences between the results for the three runs.
Although in the inner $10 \ h^{-1}$ kpc the simulations yield similar
stellar fractions, within $40 \ h^{-1}$ kpc, galactic objects in C2 have
transformed a larger fraction of the gas into stars compared with those
in C3, while galactic objects in C1 show the opposite behaviour.  
At larger radii, the mean distribution function of
stellar mass for the galactic objects in C3 is significantly larger than its
counterpart for C2, indicating that as a result of the higher cooling
efficiency, the gas in C3 has been transformed into stars further away
from the centre of mass of the objects.  Conversely, the primordial run shows 
fewer stars
in the outer region. Note that efficient energy
feedback is needed to  prevent  excessive
 star formation in the outer regions, as it  can be observed from 
 Fig.~\ref{mean-perfiles-mstar-obj}.

\begin{figure*}
\includegraphics[height=120mm]{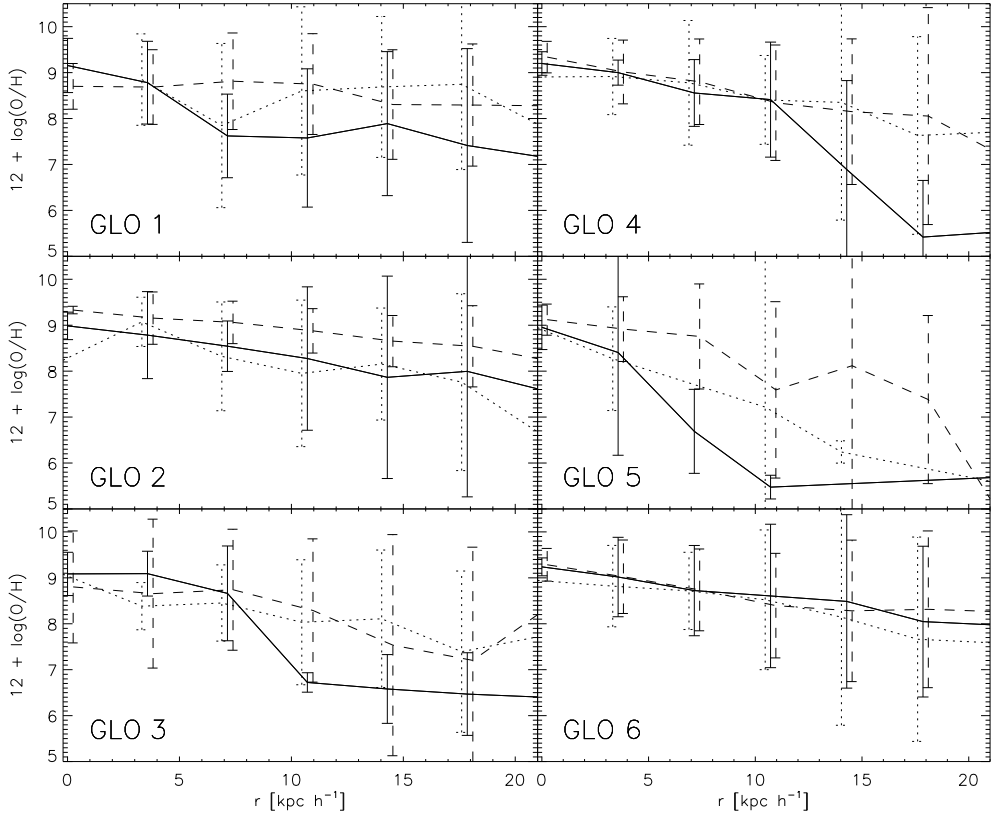}
\caption{Oxygen profiles for the gaseous components of the six
selected galactic objects from the cosmological simulations run with
primordial (dotted lines), metal-dependent (solid lines) and suprasolar (dashed lines) cooling
functions. The error bars correspond to the rms scatter of 12 + log(O/H)
around the mean.}
\label{perfiles-glos}
\end{figure*}

In order to assess how these differences affect the chemical
properties of the formed galactic objects, we analysed the metal indicator
$12+{\rm log}{\rm (O/H)}$ for the gaseous and stellar components in C1, C2
and C3. Consistent with our results for the idealized disc tests, we
found that the metallicity of the stellar components is not sensitive
to the cooling function adopted, in contrast to the gaseous components.
In Fig.~\ref{perfiles-glos}  we compare the
gaseous oxygen profiles for the galaxy samples in C1 (dotted lines), C2 (solid lines) and
C3 (dashed lines).  
Although there is a trend for systems in C3 to have a higher level
of enrichment and flatter mass-weighted abundance profiles compared
to their counterparts in C2 and C1, how much
metals affect these relations seems to depend on the particular history
of evolution of each galaxy, as can
be inferred from Figs.~\ref{sfr-glos} and ~\ref{perfiles-glos}.

\begin{figure}
\includegraphics[height=45mm]{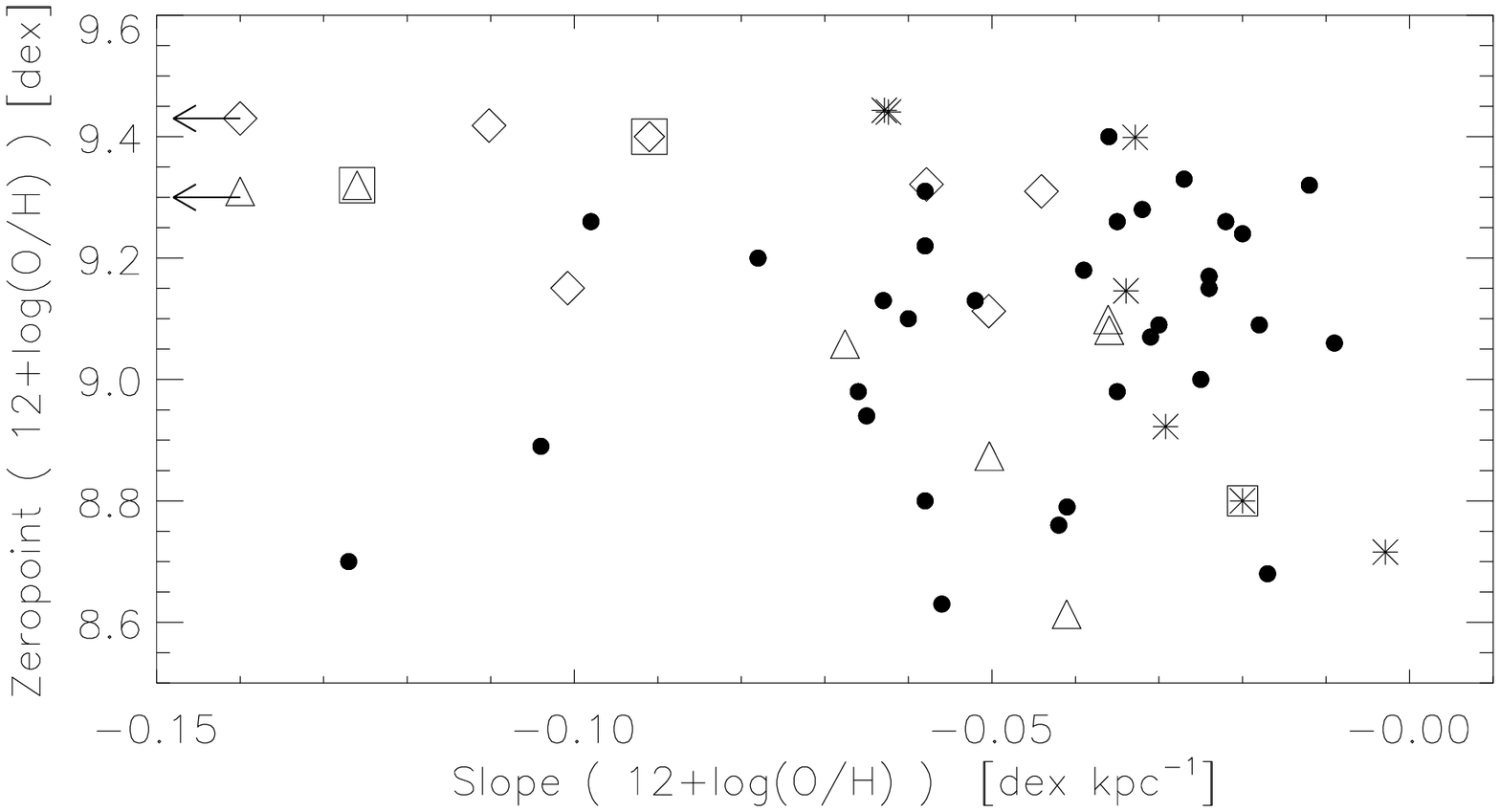}
\caption{Zero-point of the relation 12+log(O/H) for the gas components
of the selected galactic objects from the cosmological simulations as
a function of the corresponding slopes.  We have included results from
primodial (triangles), metal-dependent (diamonds) and suprasolar (asterisks) cooling
functions. For comparison, results for the tests of the isolated disc
galaxy run with primordial (squared triangle), metal-dependent (squared diamond) and suprasolar
(squared asterisk) cooling functions are also shown.  We have also
included observations from Zaritsky et al. (1994, filled circles).
The diamond and the triangle with the arrows have slopes of -0.24 and -0.23, respectively.}
\label{slope-zero}
\end{figure}

Finally, for a comparison with observational results, we performed
linear regression fits to the gaseous oxygen profiles of our simulated
galactic objects within the optical radius. By definition, the optical radius
encloses $83$ per cent of the baryonic stellar mass within the virial radius.
In Fig.~\ref{slope-zero}, we show the zero-point of these fits as a
function of the corresponding slopes for the galactic objects in C1 (triangles), 
C2 (diamonds) and C3 (asterisks).  We also include observational results from
Zaritsky et al. (1994, filled circles).  In general, the
properties of the simulated galactic objects agree reasonably well with the
range obtained in observational studies. However, the simulated
galactic objects identified in C2 tend to have steeper profiles and higher
zero-points compared with observations. Conversely, galactic objects in C3
exhibit flatter profiles.  A similar behaviour was also found in our
tests for the isolated disc galaxies run with primordial (PC, squared triangle),
metal-dependent (R2, squared diamond) and suprasolar (SC, squared asterisk) cooling
functions. The fact that these simulations do not reproduce the whole
observed range of gradients could be related to the lack of a
treatment of   energy feedback by SNe. The latter, if realistically
modelled, is expected to  trigger mass outflows  from the star-forming
regions into the halo. Since these regions are highly enriched, metals may be 
transported outwards, affecting the metallicity distributions and
flattening  the metallicity
profiles.
Our simulation run with suprasolar cooling function was
performed in order to assess the effects  of the presence of 
metals in these outer regions  on the gas cooling, mimicking 
an extreme case of metal enrichment by energy feedback.

\section{Conclusions}
\label{conclusions}

We have implemented a new model for chemical evolution within the
cosmological SPH code {\small GADGET-2}. We account for
nucleosynthesis from SNIa and SNII separately. Radiative cooling of the gas according to
self-consistent metallicity values is realized by means of precomputed
look-up tables. Our implementation was designed to allow a reasonably
accurate treatment of the main features of stellar evolution while
still not becoming computationally too expensive.

Using a number of test simulations both of isolated galaxies and of
cosmological structure formation, we have validated our implementation
and established some first basic results for the effects of chemical
enrichment. An important consequence of the presence of metals is the
increase in the cooling efficiency that it can cause, leading to
accelerated star formation. In our isolated galaxy simulations, the
SFR for test runs with metal-dependent cooling
function can be up to a factor of $1.5$ higher than the corresponding
one for a test with a primordial cooling function, and up to a factor
of $3$ lower than the one for suprasolar cooling. We found that these
differences do not originate in the gas in the central regions,
because it is already cold and dense anyway. Instead, the differences
are caused by more diffuse gas at outer radii which has yet to
cool. Here the increase of the cooling efficiency allows condensation
of gas and star formation to occur further away from the centre of the
systems.

An important implication of this result is that strong effects owing to
metal-dependent cooling can only be expected when heavy elements are
efficiently transported and mixed into diffuse gas that has yet to
cool. In our cosmological simulations, which lacked a physical model
for energy feedback, such a large-scale spreading of metals only
occurs at a low level. As perhaps  was to be expected, we
found that simulations with metal-dependent cooling produced 
similar results as simulations with a primordial cooling function. We
expect however that this could be very different if a model for
efficient metal distribution is added to our simulations. An explicit
demonstration of this has been provided by our test simulation where
we assumed suprasolar cooling rate functions, resulting in  substantially
elevated star formation activity and different abundance profiles.
Note that in this paper we explored galaxy scale systems which cool very
efficiently. Larger effects from  chemical mixing are expected in
more massive haloes such as clusters of galaxies (White \& Frenk 1991; De Lucia, Kauffmann \& White 2003).

We have also tested the impact of a number of assumptions made in our
model, and the relevance of numerical parameters.  Relaxing the
IRA for SNII has only a negligible
effect and does not change the chemical properties of the simulated
systems.  However, as also shown by previous work (e.g. Raiteri,
Villata \& Navarro 1996; Chiappini et al. 1997; Mosconi et al. 2001), the chemical
properties of both the stellar populations and the ISM
are sensitive to the time-delay associated with SNIa explosions. 
In agreement with previous results, we found that the
latter must be included in order to be able to reproduce the
observed chemical properties of galaxies.

We have shown that chemical enrichment can significantly modify the
chemical and dynamical properties of forming galactic systems.
However, the strength of the influence of metals on galaxy formation
might also depend on the physics of energy feedback.
If the latter is neglected, as we have
done here, metals are not mixed widely in gas that has yet to cool.
Effects due to metal-dependent cooling remain then  moderate. In a
forthcoming paper, we will  complement our model for chemical
enrichment with a new description of a multiphase ISM and energy
feedback, which is capable of reproducing strong outflows and regulating
the star formation activity. 
This combined model should provide a powerful tool to exploit observational data on
the chemical properties of galaxies to constrain galaxy formation.

\section*{Acknowledgements}

We thank the anonymous referee for comments that helped to improve the paper.
This work was partially supported by the European Union's ALFA-II
programme, through LENAC, the Latin American European Network for
Astrophysics and Cosmology. Simulations have been perfomed on 
Ingeld PC-Cluster funded by Fundaci\'on Antorchas.
We acknowledge support from Consejo
Nacional de Investigaciones Cient\'{\i}ficas y T\'ecnicas, Agencia de
Promoci\'on de Ciencia y Tecnolog\'{\i}a, Fundaci\'on Antorchas,
Secretar\'{\i}a de Ciencia y T\'ecnica de la Universidad Nacional de
C\'ordoba and the DAAD Exchange programme.  C.S. thanks the Alexander von Humboldt
Foundation, the Federal Ministry of Education and Research and the
Programme for Investment in the Future (ZIP) of the German Government
for partial support.


\begin{thebibliography}{}

\bibitem[ab]{ab}
Abadi M.G., Navarro J.F., Steinmetz M., Eke V.R., 2003, ApJ, 591, 499

\bibitem[Adelberger]{A03}
Adelberger K.L., Steidel C.C., Shapley A.E., Pettini M., 2003, ApJ, 584, 45

\bibitem{boi}
Boisser S., Prantzos N., 2000, MNRAS, 312, 398

\bibitem{brodie}
Brodie J.P., Huchra J.P., 1991, ApJ, 379, 157

\bibitem{bur}
Burkert A., Hensler G., 1988, A\&A, 199, 131

\bibitem{bur2}
Burkert A., Truran J.W., Hensler G., 1992, ApJ, 391, 651
 
\bibitem{chi}
Chiappini C., Matteucci F., Gratton R., 1997, ApJ, 477, 765

\bibitem{dahlem}
Dahlem M., Weaver K.A., Heckman T.M., 1998, ApJS, 118, 401

\bibitem{gabriella}
De Lucia G., Kauffmann G., White S.D.M., 2004, MNRAS, 349, 1101

\bibitem{dekel}
Dekel A., Silk J., 1986, ApJ, 303, 39

\bibitem{dts}
Dom\'{\i}nguez-Tenreiro R., Tissera P.B. \& S\'aiz A. 1998, ApJ, 508, L123

\bibitem{ettori}
Ettori S., Fabian A.C., Allen S.W., Johnstone M.N., 2002, MNRAS, 331, 635

\bibitem{fe}
Ferrini F., Matteucci F., Pardi C., Peuco U., 1992, ApJ, 387, 138

\bibitem{frye}
Frye B., Broadhurst T., Ben\'{\i}tez N., 2002, ApJ, 568, 558

\bibitem{garnett}
Garnett D.R., Shields G.A., 1987, ApJ, 317, 82

\bibitem{gin}
Gingold R.A., Monaghan J.J., 1977, MNRAS, 181, 375


\bibitem[Greggio 1996]{G96}
Greggio L., 1996, in Kunth D., Guiderdoni B., Heydari-Malayeri M.,  Thuan T., eds, The Interplay
Between Massive Star Formatin, the ISM and Galaxy Evolution. Editions Fronti$\grave{e}$res, Gif-sur-Ivette, p.89

\bibitem{GR83}
Greggio L., Renzini A., 1983, A\&A, 118, 217

\bibitem{kae}
K\"allander D., Hultman J., 1998, A\&A, 333, 399

\bibitem{kauff}
Kauffmann G., 1996, MNRAS, 281, 475

\bibitem{kch98}
Kauffmann G., Charlot S., 1998, MNRAS, 294, 705

\bibitem{ka}
Kawata D., Gibson B.K., 2003, MNRAS, 340, 908   

\bibitem{kay}
Kay S.T., Pearce F.R., Jenkins A., Frenk C.S., White S.D.M., Thomas
P.A., Couchman H.M.P., 2000, MNRAS, 316, 374


\bibitem{Kobul}
Kobulnicky H.A. et al., 2003, ApJ, 599, 1006


\bibitem{Lam}
Lamareille F., Mouhcine M., Contini T., Lewis I., Maddox S., 2004,
MNRAS, 350, 396

\bibitem{larson74}
Larson R.B., 1974, MNRAS, 169, 229

\bibitem{larson76}
Larson R.B., 1976, MNRAS, 176, 31

\bibitem{lehnert}
Lehnert M.D., Heckman T.M., 1996, ApJ, 472, 546

\bibitem[Lia et al. 2002]{Lia02}
Lia C., Portinari L., Carraro G., 2002, MNRAS, 330, 821

\bibitem{lilly}
Lilly S.J., Carollo C.M., Stockton A.N., 2003, ApJ, 597, 730

\bibitem{lucy}
Lucy L.B., 1977, ApJ, 82, 1013

\bibitem{marri}
Marri S., White S.D.M., 2003, MNRAS, 345, 561

\bibitem{martin}
Martin C.L., 2004, AAS, 205, 8901


\bibitem{metzler}
Metzler C., Evrard A., 1994, ApJ, 437, 564

\bibitem[Mosconi et al. 2001]{Mirta}
Mosconi M.B., Tissera P.B., Lambas D.G., Cora S.A., 2001, MNRAS, 325, 34 

\bibitem{mus}
Mushotzky R.F., Loewenstein M., Arnaud K., Tamura T., Fukazawa Y., Matsushita K.,
Kikuchi K., Hatsukade I., 1996, ApJ, 466, 686

\bibitem{NB91}
Navarro J.F., Benz W., 1991, ApJ, 380, 320


\bibitem{NFW96}
Navarro J.F., Frenk C.S., White S.D.M., 1996, ApJ, 462, 563

\bibitem{NFW97}
Navarro J.F., Frenk C.S., White S.D.M., 1997, ApJ, 490, 493

\bibitem{NW93}
Navarro J.F., White S.D.M., 1993, MNRAS, 265, 271


\bibitem{NW94}
Navarro J.F., White S.D.M., 1994, MNRAS, 267, 401

\bibitem{oka}
Okamoto T., Eke V.R., Frenk C.S., Jenkins A., 2005 (astro-ph/0503676)

\bibitem{pagel}
Pagel B.E.J., 1997, Nucleosynthesis and Chemical Evolution of Galaxies. Cambridge
University Press, Cambridge, chap. 8
                                                                                                           


\bibitem{pearce1}
Pearce F.R. et al. 1999, ApJ, 521, 99

\bibitem{pearce2}
Pearce F.R., Jenkins A., Frenk C.S., White S.D.M., Thomas P.A., Couchman H.M.P.,
Peacock J.A., Efstathiou G., 2001, MNRAS, 326, 649

\bibitem{proc}
Prochaska J.X., Wolfe A.M., 2002, ApJ, 566, 68

\bibitem{Raiteri}
Raiteri C.M., Villata M., Navarro J.F., 1996, A\&A, 315, 105

\bibitem{rob}
Robsertson B., Yoshida N., Springel V., Hernquist L., 2004, ApJ, 606, 32

\bibitem{rupke}
Rupke D.S., Veilleux S., Sanders D.B., 2002, ApJ, 570, 588

\bibitem{shap}
Shapley A.E., Erb D.K., Pettini M., Steidel C.C., Adelberger K.L., 2004, ApJ, 612, 122 

\bibitem{skill}
Skillman E.D., Kennicutt R.C., Hodge P.W., 1989, ApJ, 347, 875

\bibitem[Springel \& Hernquist 2002]{SH02}
Springel V., Hernquist L., 2002, MNRAS, 333, 649

\bibitem[Springel \& Hernquist 2002]{SH03}
Springel V., Hernquist L., 2003, MNRAS, 339, 289 (SH03)

\bibitem{SYW01}
Springel V., Yoshida N., White S.D.M., 2001, New Astronomy, 6, 79

\bibitem{ste}
Steinmetz M., M\"uller E., 1994, A\&A, 281, L97


\bibitem{SD93}
Sutherland R.S., Dopita M.A., 1993, ApJS, 88, 253  (SD93)

\bibitem{theis}
Theis C., Burkert A., Hensler G., 1992, A\&A, 265, 465

\bibitem{Thielemann93}
Thielemann F.K., Nomoto K., Hashimoto M., 1993, in 
Prantzos N., Vangoni-Flam E., Cass\'e N., eds, 
Origin and Evolution of the Elements. Cambridge University Press, 
Cambridge, p.299.

\bibitem{timm}
Timmes F.X., Woosley S.E.,  Weaver T.A., 1995, ApJS, 98, 617

\bibitem{tinsley79}
Tinsley B.M., Larson R.B., 1979, MNRAS, 186, 503

\bibitem{luca}
Tornatore L., Borgani S., Matteucci F., Recchi S., Tozzi P., 2004, MNRAS, 349, L19


\bibitem{Trem}
Tremonti C.A. et al., 2004, ApJ, 613, 898

\bibitem{vald}
Valdarnini R.,  2003, MNRAS, 339 , 1117

\bibitem{VDB91}
Van den Bergh S., 1991, ApJ, 369, 1

\bibitem{wef}
White S.D.M., Efstathiou G., Frenk C.S., 1993, MNRAS, 262, 1023


\bibitem{wf91}
White S.D.M., Frenk C.S., 1991, ApJ, 379, 52

\bibitem{wr78}
White S.D.M., Rees M.J., 1978, MNRAS, 183, 341

\bibitem{WW95}
Woosley S.E., Weaver T.A., 1995, ApJS, 101, 181

\bibitem{yep}
Yepes G., Kates R., Khokhlov A., Klypin A., 1997, MNRAS, 284, 235

\bibitem{Zaritsky}
Zaritsky D., Kennicutt R.C.Jr., Huchra J.P., 1994, ApJ, 420, 87



\end{thebibliography}
\end{document}